# Numerical simulations of granular dynamics II. Particle dynamics in a shaken granular material

Naomi Murdoch[†*]

Patrick Michel[*]

Derek C. Richardson[ᵛ]

Kerstin Nordstrom[ᵒ]

Christian R. Berardi[ᵒ]

Simon F. Green[†]

Wolfgang Losert[ᵒ]

† The Open University

PSSRI, Walton Hall

Milton Keynes, MK7 6AA, UK

* University of Nice Sophia Antipolis, CNRS

Observatoire de la Côte d'Azur, B.P. 4229

06304 Nice Cedex 4, France

ᵛ Department of Astronomy, University of Maryland,

College Park MD 20742-2421, USA

ᵒ Department of Physics/IPST/IREAP, University of Maryland,

College Park MD 20742-2421, USA

Available online in *Icarus* 17 March 2012

78 manuscript pages

5 tables

12 figures including 1 in colour (online version only)

Proposed running page head: Numerical simulations of shaken granular material

Please address all editorial correspondence and proofs to:

Naomi Murdoch

Laboratoire Lagrange (UMR 7293)

Observatoire de la Côte d'Azur

B.P. 4229

06304 Nice Cedex 4

France

Tel: +33 4 9200 1957

Fax: +33 4 9200 3058

E-mail: murdoch@oca.eu

# Abstract

Surfaces of planets and small bodies of our Solar System are often covered by a layer of granular material that can range from a fine regolith to a gravel-like structure of varying depths. Therefore, the dynamics of granular materials are involved in many events occurring during planetary and small-body evolution thus contributing to their geological properties.

We demonstrate that the new adaptation of the parallel $N$-body hard-sphere code `pkdgrav` has the capability to model accurately the key features of the collective motion of bidisperse granular materials in a dense regime as a result of shaking. As a stringent test of the numerical code we investigate the complex collective ordering and motion of granular material by direct comparison with laboratory experiments. We demonstrate that, as experimentally observed, the scale of the collective motion increases with increasing small-particle additive concentration.

We then extend our investigations to assess how self-gravity and external gravity affect collective motion. In our reduced-gravity simulations both the gravitational conditions and the frequency of the vibrations roughly match the conditions on asteroids subjected to seismic shaking, though real regolith is likely to be much more heterogeneous and less ordered than in our idealised simulations. We also show that collective motion can occur in a granular material under a wide range of inter-particle gravity conditions and in the absence of an external gravitational field. These investigations demonstrate the great interest of being able to simulate conditions that are to relevant planetary science yet unreachable by Earth-based laboratory experiments.

# 1. INTRODUCTION

Surfaces of planets and small bodies of our Solar System are often covered by a layer of granular material that can range from a fine regolith to a gravel-like structure of varying depths. Therefore, the dynamics of granular materials are involved in many events occurring during planetary and small-body evolution thus contributing to their geological properties.

The existence of granular material covering small bodies (asteroids, comets) was inferred from thermal infrared observations that indicate that asteroids, even small and fast rotating ones, generally have low thermal inertias compared to bare rock (Delbo and Tanga, 2009 and references therein). This granular material has also been directly observed by the two space missions that performed detailed characterizations of two near-Earth asteroids (NEAs): the NEAR-Shoemaker probe (NASA) that orbited the 30 km-size asteroid Eros for one year during 2000–2001 (Cheng et al. 2002) and the Hayabusa probe (JAXA) that visited the 500-m size asteroid Itokawa for 3 months in 2005 and successfully returned a sample from its surface to Earth in June 2010 (Tsuchiyama et al. 2011). In both cases, a layer of regolith covers the surface, although the properties of this regolith vary greatly from one object to the other. On Eros the regolith consists of fine dust of an estimated depth between 10 to 100 meters (Veverka et al. 2000). On Itokawa the regolith consists of gravel and pebbles larger than one millimeter with a depth that is probably not greater than one meter (Miyamoto et al. 2007). The reason for these different properties is not clearly understood, but we note that because of their size (mass) difference, if gravity is the discriminator, then Itokawa is expected to be as different from Eros, geologically, as Eros is from the Moon (Asphaug 2009). However, the gravity in both cases is low enough for processes such as seismic shaking to lead potentially to long-range surface motion and modification.

Seismic shaking is expected to occur when small projectiles impact the surface of small bodies on which gravity is low enough that surface motion can be influenced over long distances by seismic wave propagation. As a result of the seismic shaking, the surface granular material can be subject to various kinds of motion, among them, downslope migration and degradation, or motion leading to the erasure of small craters. Characterizing such motion under different conditions should allow us both to better assess surface evolution and to interpret observed features. For instance, there is an observed paucity of small craters on both Eros and Itokawa. Impact-induced seismic shaking, which causes the regolith to move, may erase small crater features and thus explain their paucity compared to predictions of dynamical models of projectile populations (see e.g., Richardson et al. 2004, Michel et al. 2009). The location and morphology of gravel on Itokawa indicate that the small body has experienced considerable shaking. This shaking triggered global-scale granular processes in the dry, vacuum, microgravity environment. These processes likely include landslide-like granular migration and particle sorting, which result in the segregation of fine gravels into areas of gravitational potential lows (Miyamoto et al. 2007).

Understanding the response of granular media to a variety of stresses is also important in the design of sampling tools for deployment on planetary and small-body surfaces, since the efficiency of collecting a sample from the surface of such bodies is highly dependent on its surface properties. Sample-return missions to asteroids have been studied by at least three main space agencies: ESA, JAXA (who have performed the first ever successful asteroid sample return mission) and NASA. Studies of sampling tool design generally assume that the surfaces consist of granular materials; consequently in-depth knowledge of their response to various stresses is required. While constitutive equations linking stress and strain are empirically known for most granular interactions on Earth, they involve a wide range of forces from gravity to air-grain coupling to liquid bridges due to humidity, to electrostatic effects, to surface shape and chemistry-dependent van der Waals forces. However, the inferred scaling of these equations to the gravitational and environmental conditions on other planetary bodies such as asteroids, as discussed in Scheeres et al. (2010), is currently untested.

This need to better understand granular interaction on disparate planetary bodies motivated us to conduct numerical simulations of granular material dynamics. Various numerical codes have been developed to study granular dynamics (Mehta 2007). Some of these codes are purely hydrodynamic in the sense that the granular material is represented as a fluid or as a continuum (e.g., Elaskar et al. 2000). However, the homogenization of the granular-scale physics is not necessarily appropriate and in most cases the discreteness of the particles and the forces between particles (and walls) need to be taken into account (Wada et al. 2006). Other codes, such as soft- and hard-sphere molecular dynamics codes, or codes using the Discrete Element Method (DEM) have also been developed, all of which treat the granular material as interacting solid particles (Cleary and Sawley 2002; Fraige et al. 2008; Latham et al. 2008; Szarf et al. 2011; Hong and McLennan 1992; Huilin et al. 2007; Kosinski and Hoffman 2009).

There have been several recent contributions focussing on rubble piles (defined in Richardson et al. (2002) as a special case of a gravitational aggregate) that have further highlighted the connection between granular physics and planetary science. Specifically, Walsh et al. (2008) perform hard-sphere numerical simulations to show that binary asteroids may be created by the slow spin-up of a rubble pile asteroid by means of the thermal YORP (Yarkovsky–O'Keefe–Radzievskii–Paddack) effect. The properties of binaries produced by their model match those currently observed in the small near-Earth and main-belt asteroid populations, including 1999 $KW_4$. Goldreich and Sari (2009) develop a quantitative theory for the effective dimensionless rigidity of a self-gravitating rubble pile and then use this theory to investigate the tidal evolution of rubble pile asteroids. Additionally, Sanchez and Scheeres (2011) simulate the mechanics of asteroid rubble piles using a self-gravitating soft-sphere Distinct Element Method model. One important mechanical property of such rubble piles, which is not directly included in the models, is fragility. The fragility characterises whether the material will deform in a brittle or ductile way and this may strongly influence the evolution of small bodies under external forcing.

In this paper, we employ the *N*-body hard-sphere discrete element code `pkdgrav` (Stadel 2001), which has been adapted to enable dynamic modelling of granular materials in the presence of a variety of boundary conditions (Richardson et al. 2011; hereafter Paper I). Whereas the soft-sphere methods produce contact forces by relying upon modest penetration between particles, the hard-sphere method resolves collisions by anticipating trajectory crossing. This provides hard-sphere codes with a major time advantage in the resolution of a single collision; what requires dozens of time-steps for a soft-sphere code is resolved by hard-sphere codes in a single calculation. However, both methods have their respective strengths; the hard-sphere method permits longer time-steps in the dilute regime and requires fewer material parameters, while the soft-sphere method enables more realistic treatment of friction, and is better suited to true parallelism (Schwartz et al., 2011). The advantages of the *N*-body hard-sphere adaptation of `pkdgrav` over many other hard and soft discrete element approaches include full support for parallel computation and a unique combination of the use of hierarchical tree methods to rapidly compute long-range inter-particle forces (namely gravity, when included) and to locate nearest neighbours and potential colliders (which then interact via the hard sphere potential). In addition, collisions are determined prior to advancing particle positions, ensuring that no collisions are missed and that collision circumstances are computed exactly (in general, to within the accuracy of the integration), which is a particular advantage when particles are moving rapidly. There are also options available for particle bonding to make irregular shapes that are subject to Euler's laws of rigid-body rotation with non-central impacts (cf. Richardson et al., 2009). Our approach is designed to be general and flexible: any number of walls can be combined in arbitrary ways to match the desired configuration without changing any code, whereas many existing methods are tailored for a specific geometry. Our long-term goal is to understand how scaling laws, different flow regimes, segregation, and so on change with gravity, and to apply this understanding to asteroid surfaces, without the need to simulate the surfaces in their entirety. Before we can simulate granular interactions on planetary bodies, the simulations must first be able to reproduce the dynamics of granular materials in more idealised conditions and match the results of existing laboratory experiments.

Granular material dynamics is a field of intensive research with a range of industrial applications. A variety of laboratory experiments and numerical methods have been developed to study granular dynamics, but the applications of these experiments to problems related to celestial body surfaces have only recently begun. For example, Wada et al. (2006) studied the cratering process on granular materials both by impact experiments and numerical simulations using DEM. Regolith motion resulting from seismic shaking of asteroids has also been discussed in several papers, but without simulating explicitly the dynamics of the regolith. For instance, Richardson et al. (2005) investigated the global effects of seismic activity resulting from impacts on the surface morphology of fractured asteroids. They used a Newmark slide-block analysis (Newmark, 1965), which can be applied in the regime where the regolith layer thickness is much smaller than the seismic wavelengths involved. In this case, modelling the motion of a rigid block resting on an inclined plane approximates the motion of a mobilized regolith layer. Miyamoto et al. (2007) discussed regolith migration and sorting on the asteroid Itokawa by analyzing regolith properties from images obtained by the Hayabusa probe and derived the possible regolith motion due to seismic shaking based on experiments performed on Earth.

In this paper we demonstrate that despite the fact that in our numerical scheme only two-body collisions are resolved (in comparison to the soft-sphere model where collisions between multiple particles occur), this adaptation of `pkdgrav` has the capability to accurately model the key features of the collective ordering and collective motion of a shaken granular material in a dense regime. This special case is suitable for the hard-sphere approach, but we recognise that, in general, dense systems involving multiple collisions and enduring contacts that then require a soft-sphere approach. We have in fact implemented a soft-sphere method in our code that has been tested against physical experiments (Schwartz et al. 2012, submitted), and are in the process of developing a hybrid method that switches between both approaches as needed. However, for the present purpose, the hard-sphere approach is in fact more efficient, since, as we will demonstrate (Sec. 4), the assumption of instantaneous single-point-contact collisions is appropriate for the shaking experiments and so larger time-steps can be used in the simulations.

The experiment used as a reference basis of shaking motion is described in Berardi et al. (2010). We chose this experiment as it focuses on collective behaviour where multiple particles move in a coordinated string-like fashion. We expect this collective motion will provide a more stringent test of the simulation than statistics of individual particle motion. In addition, string-like dynamics are also an indicator of fragility, an important property to consider.

We will first present the experiments that served as a reference basis for our numerical simulations of shaking motion in Sec. 2. We briefly describe the code in Sec. 3 (full details can be found in Paper I) and then Sec. 4 provides details on how the numerical simulations were performed. In Sec. 5 discussions are provided about the different analysis techniques and approaches that can be applied for such a simulation and laboratory experiment and an in-depth discussion is included highlighting the difficulties in comparing experimental and simulation data. A detailed comparison (either qualitative or, where appropriate, quantitative) is performed with the laboratory experiments. Then, in Sec. 6 we consider the consequences of varying gravitational acceleration on string frequency and length. This demonstrates the ability of our code to simulate the range of gravitational environments that can be encountered among the solid planetary bodies within our solar system. Indeed, in our reduced-gravity simulations both the gravitational conditions and the frequency of the vibrations roughly match the conditions on asteroids subjected to seismic shaking (Richardson et al., 2004; 2005), though real regolith is likely to be much more heterogeneous and less ordered than in our idealised simulations. In this same section we also demonstrate one of the unique abilities of our code: the ability to model inter-particle gravity. By removing the external gravitational field and varying the particle density we examine what happens to our granular system when the gravitational forces between the particles become increasingly strong. Finally, in Sec. 7 we discuss the relevance of our results to planetary science and discuss future perspectives.

## 2. SHAKING EXPERIMENTS

The experimental studies used as a reference in this investigation were aimed at analyzing the motion of dense configurations of bidisperse particles under vertical shaking (Berardi et al. 2010). The studies were carried out in a pseudo two-dimensional system (see Fig. 1). The goal was to analyse how a system of large and small particles arranges into ordered and disordered regions, and to elucidate the dynamics, especially in the more mobile disordered regions. The observed structure and dynamics show strong similarities to grains and grain boundaries, with large particles arranging into hexagonally ordered grain-like regions and small particles localized in grain boundaries. Additionally, the particle dynamics in the grain boundary are similar in character to a super-cooled fluid with string-like collective motion. Both the ordering and string-like dynamics are collective effects. In the experiments it was found that addition of small particles enhances the number and length of strings. Strings and ordering can both affect how fragile the ensemble of particles is, i.e., how suddenly the material jams or fails and flows under an incremental strain. Therefore by simulating this laboratory experiment we can assess the ability of the numerical code to capture collectively emerging structures and dynamics with a focus on those collective structures and dynamics that significantly affect the mechanical properties of the ensemble.

In the laboratory experiment steel spheres 3.0 mm in diameter (with the addition of 2.0 mm steel spheres, which take up from 3 to 10% of the covered surface area) were confined into a dense arrangement in a round container of 292 mm diameter with 0.1 mm separation between the large particles and a covering lid (see Fig. 1 for a diagram of the experiment and Table 1 for exact experiment conditions). When the container was vibrated vertically (at a frequency of 125 Hz with a maximum acceleration of 4.5 $g$), the dense arrangement of particles moved vertically and horizontally in a way that is characteristic for systems close to jamming. Most significantly, many of the larger (3 mm) particles formed hexagonal close-packed arrangements (the densest possible configuration of spheres in a plane). Such ordered regions, which we refer to as grains, are similar to crystalline grains in polycrystalline materials. These grains are surrounded by less densely packed, disordered regions that are named grain boundaries. It is in these grain boundary regions that most of the smaller 2 mm particles can be found. Particle motion was imaged with a high-speed high-resolution camera. From the images, the position and the motion of large and small particles were determined using an adaptation of a subpixel-accuracy particle detection and tracking algorithm (Crocker and Grier, 1996). First, images were bandpass filtered to emphasize the known particle size scale. This yields well-separated bright peaks whose positions are found with subpixel-accuracy (better than 0.13 mm) by peak-finding algorithms. To analyse motion, peaks (i.e., particles) are then tracked through image sequences that require that particles move less than half a particle diameter between frames. Comparison with the original image shows that more than 99% of all particles in each frame are detected with this algorithm. A particle track was labelled as large or small based on the average brightness of the peak. This correctly labels more than 99.9% of large particles and more than 96.1% of small particles.

The diffusion coefficient, $D$, of a system of particles has units of length-squared over time. Thus, the characteristic timescale to diffuse over a distance $L$ is $L^2/D$. The statistics of motion therefore provides a characteristic timescale when considering motion over characteristic length-scales of a particle radius. Mobile particles then can be identified based on their larger-than-expected displacement over this characteristic time interval. One characteristic of mobile particles in a system close to jamming is that mobile particles leave their "cage" of neighbours, i.e., they change neighbours. Indeed the local geometric arrangement affects mobility - mobile particles preferentially appear in grain boundaries. Similarly, string-like collective motion of mobile particles is a characteristic for systems close to jamming, particularly in glassy systems (disordered systems with extremely slow dynamics that are below and slightly above the glass transition) and dense suspensions of colloidal particles (Donati et al. 1998, Weeks et al. 2000). Rearrangements of mobile particles are characterized as strings, if particles move towards each other's previous position as if they were beads moving along a string.

The existence of this collective particle motion and the length and number of cooperatively moving clusters of particles (hereafter granular strings) can be determined (Donati et el. 1998, Aichele et al. 2003, Riggleman et al, 2006 and Zhang et al. 2009). Berardi et al. (2010) found that the surface area occupied by grain boundaries and the length and number of the granular strings increases with increasing concentrations of small particles.

Both grain boundaries and granular strings are not only useful as more subtle measures to assess whether our simulations recover collective ordering and motion, but they also affect important materials properties. Strings highlight how the yielding of granular matter is similar to the plasticity of glasses; both exhibit similar string-like collective dynamics (Stevenson and Wolynes, 2010). The presence of strings indicates that a material is fragile. In glasses, fragility is generally defined as how quickly viscosity increases when the temperature of a material is lowered toward the glass transition temperature. In granular matter, strain may be considered as the equivalent of temperature (Lui and Nagel, 1998) and fragility is associated with sudden changes in strain with increasing stress. Granular materials are, by their nature, thought to be fragile and are, therefore, prone to sudden, avalanche-like failures (Riggleman et al. 2006).

Measurements of string length offer one way to quantify this propensity for fragile behaviour – longer strings have been shown to indicate higher fragility (Dudowicz et al. 2005). In contrast, short granular strings indicate a more ductile behaviour, where failure and granular rearrangements are more uniformly distributed in space and time. String-like dynamics within grain boundaries directly affect grain boundary mobility and, therefore, play an important role in the bulk mechanical properties of more ordered systems (increased grain boundary mobility implies increased ductility) (Zhang et al. 2006).

While previous simulations have shown that strings exist in elastic disordered systems (Dudowicz et al. 2005) and in ordered systems with grain boundaries (Zhang 2006), the simulation of strings in a dissipative system in general, and in a vibrated lattice in particular has not been previously carried out. In addition, this study is the first direct comparison of the frequency and length of granular strings between experiment and simulation.

[FIGURE 1 GOES HERE]

[TABLE 1 GOES HERE]

## 3. NUMERICAL CODE

The code used for our experiments is a modified version of the cosmology code pkdgrav (Stadel 2001) that was adapted to handle hard-body collisions (Richardson et al. 2000). The granular dynamics modifications consist primarily of providing wall "primitives" to simulate the boundaries of the experimental apparatus. Implementation details are given in Paper I; only a very short summary is provided here for reference.

The code uses a second-order leapfrog scheme to integrate the equations of motion, which in this case describe ballistic trajectories in a uniform gravity field. Collision events are predicted during the linear position update portion of each integration time-step. Collisions are carried out in time order, properly accounting for repeated collisions between particles (and between particles and walls) during each step. Particles are treated as rigid spheres (so the collisions are instantaneous), with collision outcomes parameterized by coefficients of normal and tangential restitution. In addition to having a velocity vector, each particle also has a spin vector allowing particle rotation to be treated (if the tangential coefficient of restitution is < 1). The full collision resolution equations are given in Paper I.

Walls are treated as having infinite mass (so they are not affected by collisions). Paper I describes four wall primitives that can be combined in arbitrary ways: infinite plane, finite disk, infinite cylinder, and finite cylinder (the finite primitives consist of a surface combined with one or two thin rings). Certain primitives can have limited translational or rotational motion. The simulations described here used combinations of infinite planes to simulate the apparatus; two of the planes are vibrating, namely the confining lid and base-plate (Sec. 4).

# 4. NUMERICAL SIMULATION METHOD

The area covered by the particles is calculated by projecting the spheres, as 2d circles, onto the plane of the bottom of the container. The surface area coverage is then the percentage of the total container surface area that is covered by the projected surface area of all of the particles. We use this definition instead of the usual 'volume fraction' because we are dealing with a quasi-2d and not a full 3d system. Similarly, the small particle surface area coverage, or small particle concentration, is defined as the percentage of total surface area covered by the small-particle additive. In the experiments (Berardi et al. 2010) the surface area coverage studied was 85%.

The following method was used to reach a similar total surface area coverage as in experiments (Berardi et al. 2010). First, a box of 120 mm × 120 mm is constructed using the infinite plane geometries now available in `pkdgrav` (Paper I). Four infinite planes, two with normal vectors in the positive and negative *x*-directions and two with normal vectors in the positive and negative *y*-directions, form the sides of the box. One infinite plane with normal vector in the *z*-direction forms the base of the box.

The monolayer of particles at the bottom of the box is created in several steps. First, several layers of particles with radius ($R_p$) 1.5 mm are generated starting at a height of approximately 15 mm (10 $R_p$) above the base of the box (measured from the base of the box to the center of the particles in the bottom layer). The particles in each layer are evenly distributed in the *x* and *y* directions with a spacing of 4.5 mm (3 $R_p$) between the centers of the particles in each direction. The *z* position of each particle is randomly generated within the limits of ±1/$R_p$ from the mean layer height. This is to prevent all particles in each layer from impacting the base of the box simultaneously, which is unrealistic (it would be impossible in an experiment to drop all the balls from the same height exactly simultaneously). The number of particles generated depends on the desired final surface area coverage and small particle concentration. Due to the fixed inter-particle spacing there are 625 particles in each generated, and dropped, full layer. However, in order to have the correct number of particles in the box, the last layer dropped into the container is often a partial layer. The particle layers are formed one at a time by generating rows of particles from one side of the box to the other. All the particles are given a small initial velocity on the order of 1 mm s$^{-1}$ in the $x$ and $y$ directions but not in the vertical $z$ direction. Gravity (acceleration in the vertical $z$ direction) is 1 $g$ (where $g = 9.8$ m s$^{-2}$).

At the start of the simulation the generated layers of particles are allowed to fall into the box, under gravity (with no inter-particle gravity). The simulation runs until the mean vertical component of the particle translational velocity drops below a threshold of 0.1 mm s$^{-1}$, i.e., the particles are in one single layer and continue to move around the bottom of the box but are no longer bouncing in a significant way. The time-step used for the dropping phase of the simulations is such that for a particle starting from rest and falling under gravity it would take approximately 30 time-steps for it to drop one particle diameter.

A fraction of large particles are then replaced with smaller ones (radius 1 mm). In these tests the small particle concentration was a configurable parameter that was explored and thus the fraction of particles replaced depends on the desired final small particle concentration (the greater the desired final small particle concentration, the greater the number of large particles that are replaced with smaller ones). The new particles have the same position and velocity coordinates as the particles they replace but the particle radius and particle mass are updated accordingly in order to maintain a constant particle density.

A sixth infinite plane is then introduced into the simulation to provide confinement in the vertical $z$ direction (i.e., a lid is put onto the box). This infinite plane is placed parallel to the base of the box at a height of 0.1 mm above the top surface of the largest particles. This confinement allows the particles to move horizontally but prevents the particles from forming a second layer.

The four walls in the $x$ and $y$ directions are then moved inwards gradually with a speed of ~2 mm s$^{-1}$ until the box reaches a size of 100 mm × 100 mm. During this movement the box remains centered on the origin. This was found to be an effective method of increasing the surface area coverage while avoiding the problem of forming a second layer of particles. The particles are then allowed to settle in the box with all walls stationary until they reach a steady state with an equilibrated horizontal velocity.

Finally, we start the base wall and confining lid vibrating in phase. Just as in laboratory experiments (Berardi et al. 2010), the maximum acceleration of the vibration is 4.5 $g$ and the frequency is 125 Hz. The amplitude of the oscillations is thus $7.15 \times 10^{-2}$ mm. During the vibrations the downward acceleration due to gravity remains constant at 9.8 m s$^{-2}$ and there is no inter-particle gravity. For this phase of the simulations the time-step is reduced to resolve each particle-particle and particle-wall collision. During the vibration phase a particle starting from rest and falling under gravity would take approximately 130 time-steps to drop one large particle diameter.

Figure 2 shows ray-traced images of an example simulation during the vibration phase. The six infinite planes (walls) are all made completely transparent in order to facilitate observation of the particles.

Chrome steel ball bearings that were very accurate in size and shape (i.e., spheres of $3.000 \pm 0.025$ mm and $2.000 \pm 0.025$ mm diameter with an uncertainty of $10^{-3}$ mm in the particle shape) were used in the laboratory experiments, and therefore in the simulations we used an exact bimodal size distribution (i.e., spherical particles of exactly 3.0 mm and 2.0 mm diameter). The particles in the simulations have a density slightly less than the density of the experiment ball bearings (7000 kg m$^{-3}$ compared to 7900 kg m$^{-3}$). However, the particle accelerations resulting from the combination of shaking and gravity are independent of particle mass and therefore density. A normal coefficient of restitution of 0.5 (where 1.0 would mean completely elastic collisions) and a tangential coefficient of restitution of 0.9 (where 1.0 would be completely smooth), are arbitrarily chosen for all the particles in the simulation, meaning there is some dissipation of energy. These are nominal values used as an example, although a larger parameter space is explored later in Sec. 5.5. For a full list of all of the differences between the experiment and the simulations see Section 5.

The walls also have their own configurable normal and tangential coefficients of restitution. The normal coefficient of restitution is set to 0.9 and the tangential coefficient of restitution is set to 0.9 for all of the walls in these simulations. Again these values are chosen arbitrarily and a larger parameter space is explored in Sec. 5.5.

Simulations are made with 3–10% small particle surface area coverage (for a total surface area of 100 cm$^2$). The simulated time of the vibration phase is ~40 seconds. The exact conditions of the runs are given in Table 2.

As described in Section 3, the collisions in our numerical method are instantaneous. Using the equations for the duration of a binary collision provided in Campbell (2000) and the collision frequency of particles in our simulations we estimate that, during the vibration phase of the experiments, the time between collisions is at least an order of magnitude greater than the collision duration. This provides further support for our choice of numerical method, which assumes binary collisions.

[FIGURE 2 GOES HERE]

[TABLE 2 GOES HERE]

# 5. ANALYSIS AND COMPARISON WITH EXPERIMENTS

Before a detailed comparison is performed between the experimental and simulation results it is important to point out some clear differences between our numerical simulations and the original experiment of Berardi et al. (2010). Note that the experiment was performed well before the current study was defined and, therefore, its set-up was not ideally designed with the perspective of a comparison with simulations.

An extensive list of these differences is provided in Table 3 and here a few of the key differences between the experiment and simulation are highlighted. First, and possibly the most important difference, are the boundary conditions. In the experiment a large circular container is used and a rectangular area (the "test-area") of particles in the center is imaged and subsequently analysed. However, in our numerical simulations a square container was used and the motions of all the particles in the container were analysed. The circular container was not adopted in our simulations due to the difficulties involved in increasing the particle density to the correct level (cylinders with radius changing in time are not currently supported in `pkdgrav`).

Also, in the experiments the particles were not homogeneously distributed, probably due to slight inhomogeneities in shaking amplitude across the plate. This caused particles, particularly the small ones, to move closer to the edges of the field of view rather than stay homogeneously distributed across the container as in the simulations. Another factor that may have contributed to the inhomogeneity of particle distribution in the experiments is that there is a small amount of horizontal movement associated with the vertical shaking whereas in our simulations the shaking axis is constrained exactly to the vertical direction. `Pkdgrav` permits the wall vibration to be along any arbitrary axis; unfortunately, however, we cannot currently implement more than one vibration mode per wall.

Nevertheless, while there are some clear differences in the experiment and simulation, the overall dynamics of the experiment should be reproduced with the numerical simulations either qualitatively or, where appropriate, quantitatively.

[TABLE 3 GOES HERE]

## 5.1. Calculation of grains and grain boundaries

Densely packed granular systems are found to organise themselves into regions of crystallisation (grains) as well as regions of disorder (grain boundary (GB) regions) (Berardi et al. 2010).

To determine the locations of grains and regions of disorder, the simulation data is analysed using exactly the same algorithm as used for the analysis of the experiments (Berardi et al. 2010). This algorithm calculates the orientational order of the system by employing the bond orientational order parameter $\psi_6$ for each particle. The local value of $\psi_6$ is given by (Jaster 1999, Reis et al. 2006)

$$\psi_{6,i} = \left| \frac{1}{N_i} \sum_{j=1}^{N_i} e^{i6\theta_{ij}} \right|, \tag{1}$$

where $N_i$ is the number of nearest neighbours of particle $i$ and $\theta_{ij}$ is the angle between particles $i$ and $j$ and an arbitrary but fixed reference. In the analysis of the experimental and simulation data the six nearest neighbours of each particle are considered therefore, in our case, $N_i = 6$.

As a value of 1 for $\psi_6$ implies hexagonal packing, deviations of $\psi_6$ from 1 can therefore be taken as a measure of disorder. The smaller the value of $\psi_6$, the less jammed the system and thus the greater the local disorder. This measure of disorder is then used to locate the grain boundary regions. As in the experiments a value of $\psi_6 <$ 0.7 is used to define the grain boundary regions. These regions of order and disorder are a consequence of the initial particle packing in the system and not a result of the shaking behaviour.

The packing arrangements of particles during the numerical simulations are found to correctly reproduce such regions of crystallisation and regions of disorder. This is demonstrated in Fig. 3 that compares the degree of local order (i.e., $\psi_6$) at the position of each particle for both the laboratory experiment and a numerical simulation when the small particle concentration is 3%. Black signifies near-hexagonal particle packing with $\psi_6$ close to 1 while grey and white correspond to more disordered packing with $\psi_6 <$ 0.7 (i.e., GB regions). In the crystallized regions the hexagonal lattice is almost defect-free.

[FIGURE 3 GOES HERE]

It is clear that the simulations reproduce the experimentally observed collective ordering: regions of crystallisation and less ordered grain boundary regions. There are, however, some differences between the experimental and simulation particle arrangements; in the experiments the crystallized regions are randomly oriented but in the simulation the crystals are all aligned. On closer inspection it is also evident that most of the crystallized regions in our simulations are aligned from the beginning (see Fig. 4 showing the initial particle locations in the 3% and 10% small particle concentration simulations). Initially the shape of the boundaries was not considered to be important in such an investigation but it is possible that, given the rigid square boundary conditions in our simulation, our system is acting like one globally ordered single domain, which extends across the entire container. If this is the case, it could be likened to the investigations of Olafsen and Urbach (2005) who show that for spheres arranged in a hexagonal lattice at low accelerations the particle positions fluctuate continuously but no particle rearrangements are observed. As the amplitude of the acceleration is increased the spheres begin to explore all of the volume available to them and thus the density of topological defects increases dramatically. To investigate further the origin of the aligned crystals, the small-scale particle rearrangements and the effect of boundary conditions further work must be done. This is not, however, in the scope of the current paper.

[FIGURE 4 GOES HERE]

It has been experimentally observed that mean grain boundary area increases as a function of small particle concentration and therefore the amount of crystallisation decreases with increasing small particle concentration (Berardi et al. 2010). The numerical simulations have also reproduced this result. This can be demonstrated by comparing the regions of crystallisation (grains) and grain boundary regions in the 3% small particle concentration simulation in Fig. 3(b) to those of the 10% small particle concentration simulation in Fig. 5. The total area of crystallisation has decreased in Fig. 5 while the grain boundary regions have greatly increased in size. This is demonstrated more clearly in Fig. 6, which shows the trend of increasing grain boundary area with increasing small particle concentration in both the laboratory experiments and the numerical simulations. The grain boundary coverage is calculated many times over the duration of each simulation and experiment. The mean value is plotted along with the standard deviation of the mean. One difference that can be noted between the simulations and the laboratory experiment is that the fraction of the total area covered by grain boundaries is consistently greater in the simulations than in the laboratory experiments. The square boundary conditions may be partially responsible but the more probable explanation is that, as described above, in the laboratory experiments the particles (particularly the small ones) have a tendency to move to the edge of the container (due to off axis shaking and the slight inhomogeneities in shaking amplitude across the container). This means that the experimental small particle concentration in the test-area is probably not exactly the same as the small particle concentration in the entire experimental container.

It was also found that, similar to the experiments, the total grain boundary area and grain boundary locations remain almost constant in the simulations throughout the duration of the shaking and the small particles are almost all localized in grain boundary regions (see Fig. 5).

[FIGURE 5 GOES HERE]

[FIGURE 6 GOES HERE]

## 5.2. Calculation of particle velocities

In numerical simulations of granular material, instantaneous particle velocities are known at each simulation time-step thus giving a very accurate measure of the mean particle velocity at any moment in time. However, as we will explain below, since particles collide more frequently in a dense granular material than their position can be imaged, and since position measurements have inherent uncertainty, comparing the true velocity from simulations with particle velocities extracted from image sequences is not meaningful in a dense system (Xu et al., 2004). Consequently, particle velocity cannot be used as a tool to directly compare the dynamics of a simulated granular ensemble to the dynamics of a laboratory experiment. Despite this, the relative particle velocities within a simulation or experiment can be used as a diagnostic to infer further details about the location and behaviour of specific particles. This is discussed in further detail at the end of this section.

To calculate experimental particle velocities as accurately as possible, the particles are imaged at a high frame rate and the resulting particle velocities are calculated based on the particle positions in each consecutive image. However, the accuracy of the experimentally determined velocity will depend on the frame rate and resolution of the imaging, and the subsequent particle tracking. Further complications are introduced when we consider that even with very precise imaging and subsequent particle tracking there are inherently errors in any experiment. In order to extract meaningful information from the experimental data the positions of the particles must be smoothed over time. Without performing such smoothing the data would be dominated by noise. However, the more smoothing that is applied to the particle position data the more the resulting velocities are reduced. To demonstrate the effect of smoothing, Fig. 7 shows the horizontal speed of one particle in a numerical simulation over a period of 0.25 seconds. The three curves shown are for the same particle but calculated using three different methods of sampling and analysis; the first shows the instantaneous particle horizontal speed output directly from the numerical simulations, the second shows the horizontal particle speed calculated using the position coordinates of the particle sampled at 125 fps and finally the third shows the horizontal particle speed calculated using the position coordinates of the particle sampled at 125 fps and smoothed over 0.1 seconds (the technique used in the analysis of Berardi et al. (2010) experimental data). The horizontal particle speed calculated using the position coordinates of the particle sampled at 1250 fps was also calculated but is not shown because it is very similar to the instantaneous particle speed but with a smaller magnitude. The mean particle speeds listed in Table 4 for the four cases highlight even further how great the differences in the mean particle velocity can be depending on the method of sampling and analysis.

[FIGURE 7 GOES HERE]

[TABLE 4 GOES HERE]

In conclusion, as the experimental velocities can never be known exactly in a dense material where the collision rate between particles exceeds the imaging speed, or the distance between collisions is smaller than the imaging resolution, the instantaneous numerical simulation velocities can never be directly compared to experimental velocities. Even the comparison of velocities sampled at the same frame rate and analysed using the same method is unlikely to result in an accurate comparison due to inherent experimental and tracking uncertainties that do not exist in numerical simulations and that are hard to artificially introduce in simulations.

Nevertheless, the relative particle velocities within a simulation or experiment can be used as a diagnostic to infer further details about the location and behaviour of specific particles. We found, by investigating the mean particle velocities in the different regions, that there is a direct link between the particle velocity and the local ordering near the particle. The mean velocity of particles contained in the grain boundaries is much higher than the mean velocity of particles contained in the grains, as expected. Therefore the velocity of a particle in such a dense shaken system will give an indication of whether it is trapped within a crystallized grain or found within a disordered grain boundary region.

## 5.3. Calculation of long-term particle displacements

Given the difficulties involved in calculating and comparing short-term particle motion (i.e., particle velocities) a logical step is to consider the long-term particle displacements. This is particularly appropriate for the dynamical system we are trying to model given that it is only at long timescales that the complex, collaborative string-like motion should become apparent.

One method of investigating the long-term motion of particles is to consider the mean-squared displacement (MSD) profile of the system. MSD profiles are used in granular physics studies (e.g., Weeks et al., 2000 and Xu et al., 2004) but are also extensively used in many different fields of research, for example, in studies of molecular and cell biology (e.g., Sahl et al., 2010, Mika and Poolman, 2011, Fritsch and Langowski, 2010). An MSD plot indicates by what distance a particle has been displaced. The MSD is not the actual distance travelled by the particle (i.e., including all random vibrational motion) but rather the net motion in a given time (i.e., the displacement). Therefore in the long-time limit the MSD measurements focus on larger distances, where the experimental uncertainties are relatively less important. We note that based on the accuracy of the experimental particle tracking, we can assume that once we get to displacements of one pixel or more the MSDs are real and not dominated by errors.

The MSD of a system of *N* particles is calculated using the following equation:

$$MSD(\tau) = \frac{1}{N}\sum_{i=1}^{N}\left|(r_i(t+\tau) - r_i(t)\right|^2 \qquad (2)$$

where $r_i(t)$ is the position of particle *i* at time *t*, and $\tau$ is the time step between the two particle positions used to calculate the displacement.

The shape of the MSD of the shaken system is an indicator of whether the numerical simulations can successfully capture the experimentally observed complex long-term dynamics. There should be three distinct regions of the MSD profile if the dynamics are captured correctly. The MSD curve would be expected to rise initially because the particles exhibit diffusive motion as they wiggle around within the "cages" set up by their neighbours. On short timescales, the particles effectively do not "notice" the cage and simply diffuse within the cage. On longer timescales, the cage confines their motion and therefore the average displacement cannot increase with increasing measurement time. This leads to a characteristic plateau in the MSD. For long-enough times, if the system is not fully jammed, the particles will be able to break out of their cage and rearrange. This leads to a rise in the MSD curve back to a diffusive characteristic. Cage breaking involves larger-scale motion and slower dynamics that are more easily compared between experiment and simulations. Conceptually, the timescale of breaking out of the cage characterizes how far from jamming the system is. For the experimental conditions, the cage breaking time was found to be on the order of tens of seconds.

The baseline simulation curve in Fig. 8 demonstrates that the particles in our numerical simulations of this shaken system are able to reproduce the predicted dynamical evolution (i.e., the predicted MSD profile). This is the first indicator that, even though we are using a hard-sphere method that resolves only two-body interactions, we are capable of reproducing complex long-term and large-scale collective particle dynamics. The obvious differences in the magnitudes of the experimental and simulation MSD plateau values are likely to be due to pixel noise in particle position detection in the experiments. The plateau of the experimental MSD profiles (Fig. 8) is consistent with pixel noise (one pixel is equal to 0.264 mm, and thus an average change in apparent particle position of one pixel due to noise would lead to an MSD of order $(0.264 \text{ mm})^2 = 0.07 \text{ mm}^2$). This means a direct quantitative comparison of the magnitudes of the experimental and simulation MSD plateaus would be unreliable. As noted above, in the experiments the small particles have a tendency to segregate slightly near the container's walls.  By calculating the experimental MSD profile for just the large particles, and comparing it to the MSD profile for all the particles, we have determined that this slight segregation of small particles does not affect the form of the experimental MSD profiles.

Nevertheless, we can still use the MSD plots to qualitatively compare experiments and simulations by looking at the "cage-breaking" timescale. We define the cage-breaking timescale as the time at which the MSD begins to rise again. This will tell us the timescale that is needed for the jammed particles to escape their cages. Considering the MSD of the baseline simulation in Fig. 8 and comparing it to the experimental MSD it can be seen that, even though the magnitudes are different, the simulations capture the correct dynamics of the system because the cage-breaking timescale and the slope of the MSD curve are very similar for the two curves.

[FIGURE 8 GOES HERE]

## 5.4. Calculation of string-like collective motion

As previously mentioned in Sec. 5.2, the behaviour of caged motion means that a particle is always surrounded by the same neighbours. Almost fully jammed granular systems thus exhibit a characteristic timescale within which particles undergoing caged motion escape their cage (Donati et al. 1998; Aichele et al. 2003; Zhang et al. 2009). This mean time to escape defines a timescale over which string-like motions should be observable. Likewise, as described in Sec. 5.3, the same sorts of ensembles also exhibit a characteristic mean-squared displacement profile: over short (< 1 second) timescales motion is diffusive, over longer timescales (seconds) displacement is minimal and over long timescales (10s of seconds) displacement is once again diffusive. The characteristic length scale as discussed in Sec. 2 is a function of this diffusivity and the characteristic time (provided by the statistics of motion). Particles whose displacement is greater than that characteristic length scale over the characteristic time are identified as mobile particles. Some of the mobile particles are seen experimentally to exhibit cooperative motion with a string-like appearance (granular strings).

The presence of mobile particles and cooperative string-like motion in our simulations were confirmed using the same algorithms as for the experiments of Berardi et al. (2010). Briefly, we identify the timescale associated with cage-breaking, given by the peak of the non-Gaussian parameter alpha_2($t$). We calculate the actual van Hove correlation function of the system at $t^*$, $G(r,t^*)$, and compare it to a purely diffusive system. The intersection of these two curves gives a length-scale, $r^*$, which we use to identify mobile particles: particles moving a distance greater than $r^*$ in a time interval $t^*$ are mobile. From the mobile particles, we determine the subset that are also members of strings. Consider two mobile particles $i$ and $j$, at times $t$ and $t+t^*$. If particle $i$ at $t+t^*$ has moved into particle $j$'s position at $t$ (or vice versa) then the two particles are considered part of the same string. Of course, particle $i$ will rarely occupy exactly particle $j$'s original position, so we define a cut-off distance, $\Delta_c$, which is how close particle $i$ has to be to particle $j$'s position. For historical reasons, we choose this $\Delta_c$ to be 0.6 times a large particle diameter (Donati et al. 1998). It should be noted that the absolute values of the average number of strings and string length are dependent on one's choice of the value of $\Delta_c$. However, the overall *trend* in these parameters with respect to particle concentration is insensitive to the exact choice of $\Delta_c$.

As in the experiments, the granular strings were located in the grain boundary regions and not within the grains. The number of granular strings at a given time during the simulations and the average string length (in number of particles) were also analysed for each small particle concentration (see Figs. 9 and 10). It was experimentally observed that as the concentration of small particles is increased, the scale of the cooperative motion also increases, i.e., the number and size of the granular strings increase with increasing small particle concentration (Berardi et al. 2010). The numerical simulations reproduce the correct dependence on small particle concentration, with both the number and length of granular strings increasing with small particle concentration.

As the experimental and simulated systems were not the same size and thus do not contain the same number of particles it is not possible to quantitatively compare the average number of granular strings detected per time-step. Nevertheless, as shown in Fig. 9, the average number of granular strings detected in the numerical simulations does show the correct dependence on small particle concentration, i.e., the larger the concentration of small particles, the more granular strings are detected. We can, however, perform an accurate quantitative comparison investigating the average size of the granular strings in the system and how this depends on small particle concentration. It is demonstrated, in Fig. 10, that not only do the numerical simulation results show the correct dependence of granular string length on small particle concentration but also we reproduce almost exactly the experimental results. The slight discrepancy between simulation and experimental results at 3% small particle concentration may be due to experimental uncertainties that are not taken into account here. This final test demonstrates clearly the capabilities of this adaptation of `pkdgrav` to accurately model the key features of the collective ordering and motion of a shaken granular material in a dense regime.

[FIGURE 9 GOES HERE]

[FIGURE 10 GOES HERE]

## 5.5. Sensitivity to simulation parameters

One of the clear advantages of numerical simulations over experiments is the ability to investigate a much wider parameter space, often including environmental conditions or material properties that are not easily investigated experimentally. Conversely, one of the key drawbacks of numerical simulations is the capacity to "tune parameters" to sometimes unrealistic values in order to match the desired outcomes.

Here we present the results of several investigations performed into the sensitivity of the simulations to the internal parameters, with two aims: firstly to develop our understanding of how the simulation parameters influence the particle behaviour in such a system, and secondly to allow us to conclude beyond any doubt that `pkdgrav` can accurately model the correct behaviour and physics of granular materials in a dense regime as a result of shaking and that the results are not random and are in fact closely related to the initial conditions.

Several investigations were performed varying the key internal parameters of the numerical simulations. Given the volume of tests performed, only the important results and trends will be discussed in detail; however, details of all the simulations performed can be found in Table 5. In each of these tests the simulation set-up was identical to the one described in Sec. 4. The surface area, surface area coverage and initial particle packing configurations were unchanged.

The first set of investigations focussed on the influence of changing the coefficients of restitution of the particles and the walls with particular attention paid to the resulting mean particle velocities and MSD profiles as a way of interpreting the system behaviour. As discussed previously there are two types of coefficient of restitution: the normal coefficient of restitution (where 1.0 would mean completely elastic collisions) and the tangential coefficient of restitution (where 1.0 would be completely smooth surfaces). By changing these parameters we can produce a system where there is a varying degree of energy dissipation and coupling between particles and particles and walls. We note that for large coefficients of restitution, particularly the normal coefficient of restitution of the particles, there is an increase in the MSD profile only at very long timescales and thus the particles, while losing less energy in each collision, need a longer time to break out of their local cage. Conversely, low coefficients of restitution result in particles breaking out of their cage rapidly (see Fig. 8). From this we can conclude that lower coefficients of restitution, i.e., lower particle velocities and more inter-particle and inter-wall coupling, are likely to result in more complex cooperative motion despite the fact that the low coefficients also reduce the overall energy of the system.

A separate investigation considered the effect of changing the simulation time-step on the system dynamics. The simulation time-step was changed so that for a large particle starting from rest and falling under Earth's gravity it would take 130, 217, 650, 1300 and 6500 time-steps to fall one particle diameter. The data output frequency was kept constant at 125 Hz. It was found that although the mean particle velocities remain unchanged at all times during the simulation regardless of the time-step, we note a certain small variation in the MSD. However, the overall MSD trends are consistent and qualitatively similar. Additionally, there is no trend in the variations of the MSD profiles with decreasing time-step; as the simulation time-step decreases we are not converging to a more accurate MSD profile. Therefore, we can conclude that the variations in MSD profiles, caused by changing the time-step, are most probably random and, as long as the time-step is small enough, they are not critical.

The final investigation of the simulation parameters considered the impact of changing the simulation output frequency, or the rate at which the data is sampled, while keeping the time-step of the numerical simulations constant. The data were sampled at 125, 250, 417, 650 and 1250 Hz. The resulting MSD profiles were largely unaffected by the changes in sampling frequency and thus show no sampling bias. However, as discussed in Sec. 5.2, the data sampling frequency can have a large influence on the measured velocities of the system. It must, however, be noted that simulations are "perfect" and don't contain inherent real world characteristics so these types of biases are much more evident than they would be in laboratory experiments. Nevertheless, as the sampling frequency can be chosen with much greater flexibility in numerical simulations, such simulations could be an invaluable tool to help experimentalists determine what level of sampling frequency is necessary to avoid any potential sampling biases.

[TABLE 5 GOES HERE]

# 6. PLACING THE SIMULATIONS IN A VARYING GRAVITATIONAL CONTEXT

Finally, we apply our measure of collective dynamics and fragility – string length and number – to simulate conditions that are hard to replicate experimentally. First, we consider the consequences of varying the external gravity on string frequency and length. Next, we demonstrate one of the unique abilities of our code: the ability to model inter-particle gravity. By varying the particle density we examine what happens to our granular system when the gravitational forces between the particles become increasingly strong.

## 6.1 Varying the external gravitational acceleration

In this section, we consider the consequences of varying the external gravitational acceleration on string frequency and length. This demonstrates the ability of our code to simulate the range of gravitational environments that can be encountered among the solid planetary bodies within our solar system. The external gravity is varied from $0.01 - 10$ *g*. The particle density remains unchanged and the vibrational amplitude and frequency remain the same as in the experiments and baseline simulations. In addition, each simulation has an identical initial configuration (i.e., identical initial particle locations).

In our reduced-gravity simulations (when *g* << 1) the gravitational acceleration is of a magnitude similar to that found on the surfaces of asteroids. In addition, the frequency of the vibrations roughly matches the conditions on asteroids subjected to seismic shaking (Richardson et al., 2004; 2005). Although we are aware that vibrations due to seismic shaking on an asteroid are not likely to act always in the same direction we can still expect string-like collective motion in excited, heterogeneously sized and shaped regolith. We do not, however, expect grain boundaries to occur in regolith. In our ordered and idealised system of equal sized spheres the grain boundaries are the heterogeneous regions where collective rearrangements take place. In glassy i.e., disordered systems, string-like motion is expected to occur everywhere. Nevertheless, we have used our simulations to demonstrate that we have the capability of varying the external gravitational acceleration and to show the sensitivity of our idealised system to such variations in the external gravitational acceleration.

We find that the length of strings and the frequency of strings decrease with increasing external gravitational acceleration, as shown in Fig. 11. This indicates that decreasing the external gravitational acceleration makes the granular ensemble more fragile when subjected to local excitation amplitudes. At the same time, cooler, less energetic systems appear to become less fragile.

[FIGURE 11 GOES HERE]

## 6.2 Varying the inter-particle gravitational acceleration

In this section we demonstrate one of the unique abilities of our code: the ability to model inter-particle gravity. By varying the particle density over several orders of magnitude we examine what happens to our granular system when the gravitational forces between the particles become increasingly strong. In this investigation the external gravitational field was removed (i.e., the system is in 'zero-gravity') and the vibrational amplitude and frequency remain the same as in the experiments and baseline simulations. Again, each simulation has an identical initial configuration (i.e., identical initial particle locations).

The measured resulting changes in string properties with varying particle density, and thus varying inter-particle gravity, are shown in Fig. 12. Our system has a natural length-scale, which is the distance between particles at which the gravitational potential energy between two particles is equal to the mean particle kinetic energy. For each simulation we have determined this natural length-scale for both the large and the small particles. We find that, for the largest particles (which are the most numerous), the natural length-scale is equal to one particle diameter for particles of density ~1.5 x $10^{13}$ kg m$^{-3}$ (for the smaller particles, the density giving a natural length-scale of approximately one particle diameter is slightly larger). This indicates that, at the smaller densities we have tested, the dynamics of the system are dominated by the kinetic energy of the particles and at the largest densities the gravitational potential energy may begin to play an important role in the dynamics of the system. Our analysis and simulations indicate that the scale of the collective motion decreases in the region where the gravitational potential energy between particles is of a comparable magnitude to the mean particle kinetic energy (see Fig. 12). This may be because the inter-particle gravity is acting like an adhesive force between particles thus reducing the fragility of the system. A full study would be needed to confirm this, but this is outside the scope of this paper.

We note that the densities considered for this investigation are unrealistically large. However, the pair-wise gravitational attraction between two identical particles in contact, $F_g \propto \rho^2 r^4$, where $\rho$ is the bulk density of the particles and $r$ is the radius. Therefore, our study varying density is equivalent to a study where the radius of the particles of standard density of $10^3$ kg m$^{-3}$ is varied from ~1 mm to ~400 m.

Finally, we note that in future studies it may be useful to consider a full 3d system to investigate in more detail the role that self-gravity plays in affecting collective motion.

[FIGURE 12 GOES HERE]

# 7. DISCUSSION, RELEVANCE TO PLANETARY SCIENCE AND FUTURE WORK

We have demonstrated that the implementation of the hard-sphere discrete element method in the *N*-body code `pkdgrav` is capable of simulating the key features of the complex collective motion of a particular densely packed, driven granular system. While there are some clear differences in the experiment and simulation, the overall dynamics of the experiment have been reproduced either qualitatively or, where appropriate, quantitatively.

As a first test we showed that our numerical simulations correctly reproduce the regions of crystallisation (grains) and regions of disorder (grain boundaries) found experimentally. We discussed the difficulties involved when trying to compare experiments and simulations quantitatively and concluded that due to inherent experimental and tracking errors, particle velocities are not a meaningful variable to compare. This is particularly true in a dense material where the collision rate between particles exceeds the imaging speed, and the distance between collisions is smaller than the imaging resolution. We suggested that mean-squared displacement (MSD) profiles are a more reliable means of comparison and have matched the experimental cage-breaking timescale (i.e., the timescale at which jammed particles may escape their cage) and gradient of the subsequent rise of the MSD profile. As a final test we examined our system for mobile particles and string-like collective motion. In previous studies such string-like motion has been found to be a signature of fragility; an important material property that indicates how quickly a material softens under increasing external forcing. We found that mobile particles are present and that our numerical simulations reproduce the key features of the experimentally observed string-like collective motion of such mobile particles even though the simulations are based on pairwise collisions only.  Just as in the experiments, we demonstrated that the scale and frequency of occurrence of the collective motion of the shaken granular system can be increased by the addition of small particles. The close match found between experimental and simulation results during a quantitative comparison of the average size of the granular strings is further validation of our numerical scheme. We also successfully demonstrated that in this dense regime the behaviour and physics of the shaken granular matter predicted by our numerical simulations are not random and are closely related to the particle parameters and simulation initial conditions.

As mentioned above and discussed in detail in Sec. 2, previous studies have shown that the presence of granular strings indicates that a material is fragile, i.e., prone to more sudden, avalanche-like failures. However, short granular strings indicate a more ductile behaviour. Flow of granular material has been inferred from observations of the asteroid Itokawa's surface taken by the Hayabusa spacecraft and from observations of the asteroid Lutetia's surface taken by the Rosetta spacecraft.  As noted by Miyamoto et al. (2007), there are strong indications that gravels on Itokawa, based on their locations and morphological characteristics on the surface, were relocated after their accumulation/deposition, implying that the surface has been subject to global vibrations. These vibrations are likely to have triggered global-scale granular processes including landslide-like granular flows and particle sorting that result in the segregation of the fine gravels into areas of potential lows.  However, from the existing observations, one cannot easily discriminate between gradual and abrupt changes on Itokawa's surface. From a strictly mechanical point of view, we may expect some differences between these two modes of migration. Simulations in conditions close to the asteroid environment are required to understand what these differences could be and which circumstances are necessary to lead to the observed characteristics. Such understanding is crucial for interpreting observations of asteroid surfaces, and to derive the regolith properties. For instance, the presence of granular strings could be one possible explanation for observed changes, if we were able to assess that they occurred suddenly.

Evidently these problems cannot fully be investigated by the hard-sphere approach alone and probably the best strategy will be to use a hybrid method that selects the appropriate approach (dilute vs. dense regime) as needed. Indeed, to determine the validity of the hard-sphere approach, we would need to understand the flow rate for asteroidal material at which binary collisions may be dominant based on particle hardness, and so on. However, these properties are not yet known and investigating the full parameter space is beyond the scope of this paper. The present study aims to evaluate one component of the overall approach, namely the hard-sphere discrete element method. Here we demonstrate that the implementation in `pkdgrav` of this approach, which is valid for dilute regimes (e.g., planetary rings), is capable of reproducing the dynamical behavior of a specific dense system as well. The next step will be to validate the implementation of the soft-sphere approach (Schwartz et al., 2012) and eventually develop a hybrid method that includes both approaches. For example, study of granular avalanches on asteroid surfaces could benefit from such hybridisation, since in an avalanche flow, the particles close to the top of falling particle layer, which undergo many collisions, exhibit fast, dilute flows, while particles at the bottom are in slow, dense flows, i.e., particles remain in contact with neighbours for long intervals. In the meantime, the code's ability to create arbitrary shapes through particle bonding, and to vary the external gravity, can still be exploited with the hard-sphere approach to investigate particle motions in dilute environments. Taking again the example of a granular avalanche, the dilute layer could easily be studied with our hard-sphere approach, as Kharhar et al. (1997) did by separating the granular material in a rotating drum into a "rapid-flow region" and a fixed bed. We could also study the evolution of ejecta from impacts or the fluidisation of energised regolith on small-body surfaces, such as regolith motion resulting from the sudden impact of a small projectile on an asteroid's surface, as well as particle ring dynamics (e.g., Perrine et al. 2011), tidal encounters (for which the encounter time is short or comparable to the dynamical time; e.g., Richardson et al., 1998), and so on.

Finally, having the number and length of strings as a metric of collective motion and fragility allows us to take full advantage of the possibilities that are opened up by the simplicity of hard sphere simulations: we can include self-gravity, vary external gravity, and explore their effect on the indicators of fragility of the granular material. We find that external gravity changes collective behaviour: ensembles of particles exhibit more collective motion and, therefore, appear more fragile when held in place by lower external gravity. The fragility of planetary bodies is particularly important as it is potentially related to the onset of sudden fracture or failure events of the body. Our simulations varying the inter-particle gravity of our system suggest that collective motion and thus, fragility, may depend closely on the balance between the gravitational potential energy and the kinetic energy of the system. This interesting discovery is highly relevant for small bodies and would be very interesting to consider in future studies.

Currently this work considers only a quasi-2d system. However, the advantage of our numerical code is that we can easily extend this to a fully 3d system.  As our analysis indicates that collective behaviour is correctly captured in the simulations, string lengths and numbers may also be measured in 3d as metrics for fragility. We emphasise that this present study is a first step that is necessary to ensure that the physics involved in the system we investigated is well computed and that gives us confidence that we can accomplish the next steps, which are to apply it directly to actual planetary science problems. Using our numerical simulations would allow us to perform investigations that are difficult to access with experimental observations such as investigating collective motion within the bulk of a 3d granular system or investigating the fragility of bodies to deformations due to, e.g., tidal forces. The role of rotational versus translational motion in driving string formation is also an interesting physics question, which could be addressed with our simulations but is outside of the scope of this current paper.   Additionally, the influence of boundary conditions, the tangential coefficient of friction and also cohesion on the formation of granular strings will be explored in future work.

**ACKNOWLEDGMENTS**

We would like to thank The Open University, Thales Alenia Space and the French National Program of Planetology for financial support. DCR acknowledges support from the National Aeronautics and Space Administration under Grant No. NNX08AM39G issued through the Office of Space Science. WL was supported by NSF-DMR0907146. Some simulations in this work were performed on the CRIMSON Beowulf cluster at OCA. Ray tracing for Fig. 2 was carried out using the Persistence of Vision Raytracer[1]. This study was done as part of the International Team collaboration number 202 sponsored by the International Space Science Institute (ISSI) in Switzerland. We would also like to thank two anonymous reviewers for their very constructive comments.

---

[1] http://www.povray.org/

TABLE 1. The experimental conditions. The entire experiment was 670 cm$^2$ in area, although during the analysis a test area of only 230.11 cm$^2$ was considered (see Sec. 5 for more details and discussion on the differences between experiment and simulations). An estimation was made of the worst-case error in the total number of particles in the entire experiment. From this worst-case error, the maximum possible uncertainties in total surface area, small particle surface area coverage and the total number of particles in the test area were calculated.

| Small particle surface area coverage (% of total covered surface area) | Total surface area coverage (%) | Total number of particles in test area |
|---|---|---|
| 3 ± 0.1 | 85 ± 0.21 | 2615 ± 20 |
| 5 ± 0.1 | 85 ± 0.21 | 2600 ± 20 |
| 7 ± 0.1 | 85 ± 0.21 | 2640 ± 20 |
| 10 ± 0.1 | 85 ± 0.21 | 2680 ± 20 |

TABLE 2. The conditions of the numerical simulations. The normal coefficient of restitution of the particles is 0.5 (where 1.0 would mean completely elastic collisions) and the tangential coefficient of restitution is 0.9 (where 1.0 would be completely smooth).

| Small particle surface area coverage (% of total covered surface area) | Total surface area coverage (%) | Total number of particles |
|---|---|---|
| 2.91 | 83.76 | 1247 |
| 4.84 | 83.85 | 1277 |
| 6.89 | 83.83 | 1307 |
| 10.16 | 83.63 | 1352 |

TABLE 3. Detailed list of all differences between experiments and simulations.

| PARAMETER | EXPERIMENT | SIMULATION |
|---|---|---|
| **Total surface area** | 670.1 cm$^2$ | 100.0 cm$^2$ |
| **Surface area considered in analysis** | 214.6 cm$^2$ | 100.0 cm$^2$ |
| **Shape of container** | Circular (292 mm diameter) | Square (100 mm x 100 mm) |
| **Shape of area considered in analysis** | Rectangular (169 mm x 127 mm) | Square (100 mm x 100 mm) |
| **Shaking axis** | In the z-direction with a small acceleration (< 0.5%) in the horizontal (x-y) plane | In the z-direction |
| **Particle sizes** | Diameter tolerance of 0.025 mm | Exactly 2.0 mm or 3.0 mm diameter |
| **Particle density** | 7900 kg m$^{-3}$ | 7000 kg m$^{-3}$ |
| **Particle shapes** | Uncertainty of $10^{-6}$ m | Exactly spherical |
| **Small particle concentration** | Calculated by weight for the entire container | Calculated exactly |
| **Movement of walls during shaking** | Entire container shaken so all container walls shaken in phase | Top and bottom walls shake in phase, side walls are stationary |
| **Bottom wall shape** | Slightly convex | Exactly planar |

TABLE 4. The mean horizontal speed of one particle in a numerical simulation during a period of 5 seconds calculated using four different methods.

| METHOD | Mean horizontal particle speed (mm s$^{-1}$) |
|---|---|
| A: Instantaneous speed | $9.9 \pm 5.7$ |
| B: 1250 fps sampling | $8.5 \pm 4.9$ |
| C: 125 fps sampling | $4.4 \pm 2.4$ |
| D: 125 fps sampling with smoothing | $2.0 \pm 1.1$ |

TABLE 5. The different simulations performed over the course of this study. The parameters in bold are the baseline parameters used in this study.

| Simulation time-step (number of time-steps to fall $1d$ under gravity) | Normal coefficient of restitution (particles) | Tangential coefficient of restitution (particles) | Normal coefficient of restitution (walls) | Tangential coefficient of restitution (walls) |
|---|---|---|---|---|
| 130 | 0.1 | 0.1 | 0.1 | 0.1 |
| 130 | 0.5 | 0.5 | 0.5 | 0.5 |
| 130 | 0.9 | 0.9 | 0.9 | 0.9 |
| 130 | 0.5 | 0.1 | 0.1 | 0.1 |
| 130 | 0.9 | 0.1 | 0.1 | 0.1 |
| 130 | 0.1 | 0.5 | 0.5 | 0.5 |
| 130 | 1.0 | 0.5 | 0.5 | 0.5 |
| 130 | 0.1 | 0.9 | 0.9 | 0.9 |
| **130** | **0.5** | **0.9** | **0.9** | **0.9** |
| 130 | 0.1 | 0.5 | 0.1 | 0.1 |
| 130 | 0.1 | 0.9 | 0.1 | 0.1 |
| 130 | 0.5 | 0.1 | 0.5 | 0.5 |
| 130 | 0.5 | 1.0 | 0.5 | 0.5 |
| 130 | 0.9 | 0.1 | 0.9 | 0.9 |
| 130 | 0.9 | 0.5 | 0.9 | 0.9 |
| 130 | 0.1 | 0.1 | 0.5 | 0.1 |
| 130 | 0.1 | 0.1 | 0.9 | 0.1 |
| 130 | 0.5 | 0.5 | 0.1 | 0.5 |
| 130 | 0.5 | 0.5 | 1.0 | 0.5 |
| 130 | 0.9 | 0.9 | 0.1 | 0.9 |
| 130 | 0.9 | 0.9 | 0.5 | 0.9 |
| 130 | 0.1 | 0.1 | 0.1 | 0.5 |
| 130 | 0.1 | 0.1 | 0.1 | 0.9 |
| 130 | 0.5 | 0.5 | 0.5 | 0.1 |
| 130 | 0.5 | 0.5 | 0.5 | 1.0 |
| 130 | 0.9 | 0.9 | 0.9 | 0.1 |
| 130 | 0.9 | 0.9 | 0.9 | 0.5 |
| 130 | 0.4 | 0.4 | 0.6 | 0.6 |
| 130 | 0.3 | 0.9 | 0.7 | 0.95 |
| 130 | 0.3 | 0.6 | 0.5 | 0.5 |
| 130 | 0.4 | 0.8 | 0.3 | 0.95 |
| 130 | 0.8 | 0.9 | 0.6 | 0.95 |
| 217 | 0.8 | 0.9 | 0.6 | 0.95 |
| 650 | 0.8 | 0.9 | 0.6 | 0.95 |
| 1300 | 0.8 | 0.9 | 0.6 | 0.95 |
| 6500 | 0.8 | 0.9 | 0.6 | 0.95 |
| 130 | 0.4 | 0.4 | 0.6 | 0.6 |
| 130 | 0.3 | 0.9 | 0.7 | 0.95 |
| 130 | 0.3 | 0.6 | 0.5 | 0.5 |
| 130 | 0.4 | 0.8 | 0.3 | 0.95 |

# Figure Captions

**Figure 1.** Schematic of the experiment used in Berardi et al. (2010). The container depth is 3 mm and the separation between the top of the largest particles and the confining lid is 0.1 mm. The base plate, container and the confining lid vibrate together at a frequency of 125 Hz and with a maximum acceleration of 4.5 $g$. Note: Figure not to scale.

**Figure 2.** Ray-traced images of a simulation during the vibration phase as seen from (a) the side and (b) above. Particles are contained in a box of 100 mm × 100 mm. There is a confining lid in the $z$-direction 0.1 mm above the largest particles (see Fig 1). All walls are made transparent to facilitate observation. There are two sizes of particles: 1.5 mm radius (red) and 1 mm radius (yellow). The total surface area coverage is 83.63%. The 1 mm radius particles cover 8.50% of the entire container surface area (i.e., 8.50% of 100 mm$^2$), which is equivalent to 10.16% of the total covered surface area (i.e., 10.16% of 83.63 mm$^2$).

**Figure 3.** The degree of local order (i.e., packing density) at the position of each particle when the small particle concentration is 3%. Results are shown for both the laboratory experiment (a) and a numerical simulation (b). Grain Boundary (GB) regions determined using the algorithm of Berardi et al. (2010). Black signifies near-hexagonal particle packing with $\psi_6$ close to 1. Grey and white correspond to more disordered packing with $\psi_6 < 0.7$ (i.e., GB regions). See Equation (1) for the definition of $\psi_6$. Particles are not drawn to scale.

**Figure 4.** The locations of all particles at the start of the vibration phase of the numerical simulations when the small particle concentration is (a) 3% and (b) 10%. Large (3mm) particles are white, small (2mm) particles are black. Figures are drawn to scale. In both figures regions of crystallisation can be seen.

**Figure 5.** The degree of local order (i.e., packing density) at the position of each particle in a numerical simulation when the small particle concentration is 10%. Grain Boundary (GB) regions determined using the algorithm of Berardi et al. (2010). Black signifies near-hexagonal particle packing with $\psi_6$ close to 1. Grey and white correspond to more disordered packing with $\psi_6 < 0.7$ (i.e., GB regions). See Equation (1) for the definition of $\psi_6$. The locations of the small (2 mm) particles are all marked with an X. The laboratory experiment results are not shown here, but in both the experiments and the numerical simulations the small particles are almost all located in grain boundaries. Particles are not drawn to scale.

**Figure 6.** Grain boundary coverage, measured as the percentage of covered surface in grain boundaries, as a function of small particle concentration for the laboratory experiment (open circles) and the baseline numerical simulation (filled circles). The grain boundary coverage is calculated many times (approximately once per second) over the duration of each simulation and experiment and the mean value is plotted along with the standard deviation of the mean. Increasing grain boundary area with increasing small particle concentration can clearly be seen for both data sets. The quantitative differences are discussed in Sec. 5.

**Figure 7.** The speed of one particle in a numerical simulation over a period of 0.25 seconds, calculated using three different methods: (white diamonds): the instantaneous particle speed calculated during the numerical simulations; (black dots): using the coordinates of the particle sampled at 125fps; (open white circles): using the coordinates of the particle sampled at 125 fps and smoothed over 0.1 seconds.

**Figure 8.** The mean-square displacement (MSD) curves are shown for the laboratory experiment with 10% small particle concentration (dotted line) and three different numerical simulations also with 10% small particle concentration. The similar shapes of the experimental and simulation MSD curves, and the magnitude discrepancy, are discussed in Section 5. Also shown for comparison, and to demonstrate the sensitivity of our simulations to different parameters (see Sec 5.5), are two extreme 10% small particle concentration numerical simulation MSD profiles; case A in which the particles break out of their cage very early (dashed line) and case B where the particles remained jammed over much longer timescales (dashed - dotted line). The parameters for cases A and B are, respectively: normal coefficient of restitution of the particles = 0.1 and 0.9; tangential coefficient of restitution of the particles = 0.5 and 0.9; normal coefficient of restitution of the walls = 0.5 and 0.9; tangential coefficient of restitution of the walls = 0.5 and 0.9.

**Figure 9.** Average number of granular strings at a given time-step as a function of additive concentration for the baseline simulation, defined as the percentage of total area covered by the small-particle additive. The errors bars are the standard error of the mean for one experiment at one small particle concentration. As each experiment

is thousands of frames, we expect to sample all configurations and so report the standard error and not the standard deviation.

**Figure 10.** Average granular string length in particles as a function of small particle surface area coverage, defined as the percentage of total surface area covered by the small-particle additive from the laboratory experiment (open circles) and the numerical simulations (filled circles). The error bars, which are smaller than the size of the markers, are the standard error of the mean for one experiment at one small particle concentration.. As each experiment measures thousands of strings, we expect to sample all configurations and so report the standard error and not the standard deviation.

**Figure 11.** (a) Average number of granular strings and (b) average granular string length at a given time-step as a function of varying external gravity for simulations with 10% small particle concentration. The errors bars, which are sometimes smaller than the markers, are the standard error of the mean for one experiment at one small particle concentration. As each experiment is thousands of frames, we expect to sample all configurations and so report the standard error and not the standard deviation.

**Figure 12.** (a) Average number of granular strings and (b) average granular string length at a given time-step as a function of varying particle density for simulations with 10% small particle concentration and inter-particle gravity but no external gravity. The errors bars, which are sometimes smaller than the markers, are the standard error of the mean for one experiment at one small particle concentration. As each experiment is thousands of frames, we expect to sample all configurations and so

report the standard error and not the standard deviation.

FIGURE 1

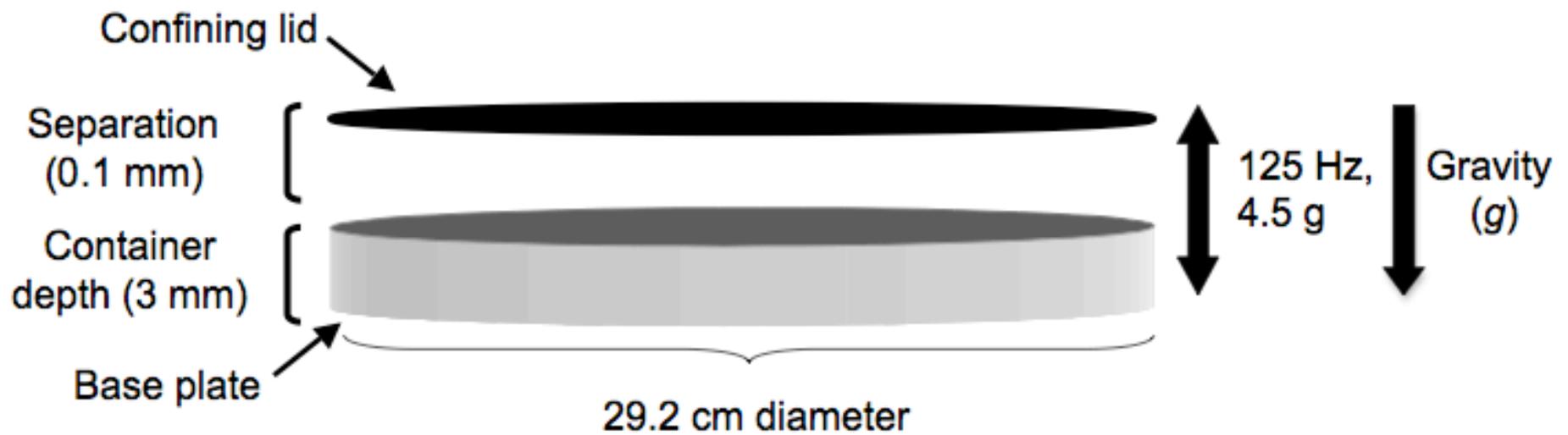

FIGURE 2

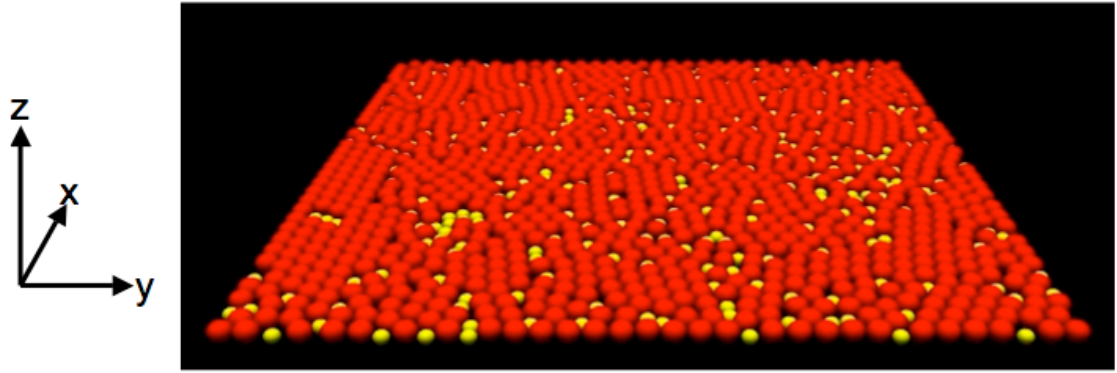

(a)

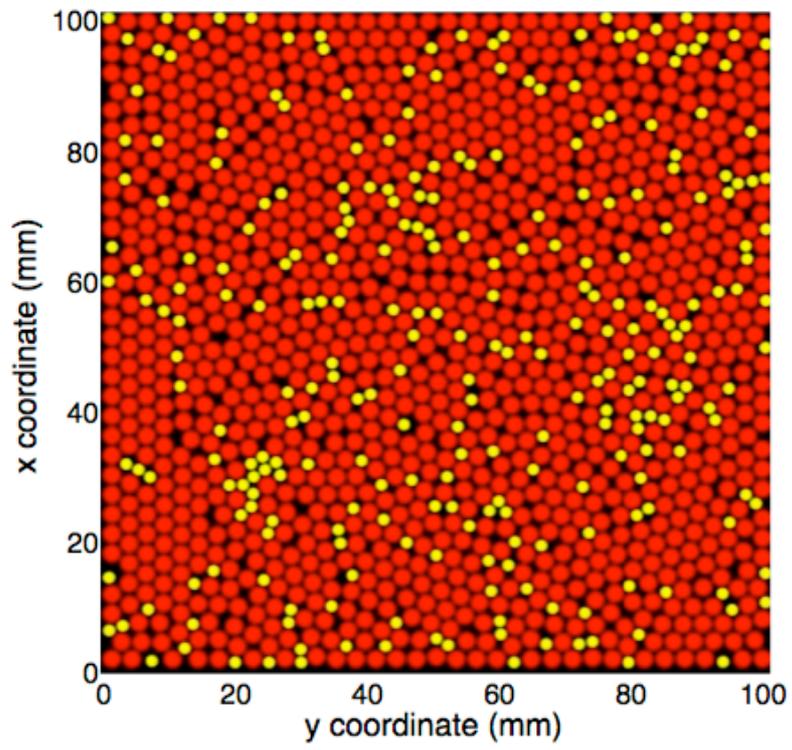

(b)

FIGURE 3

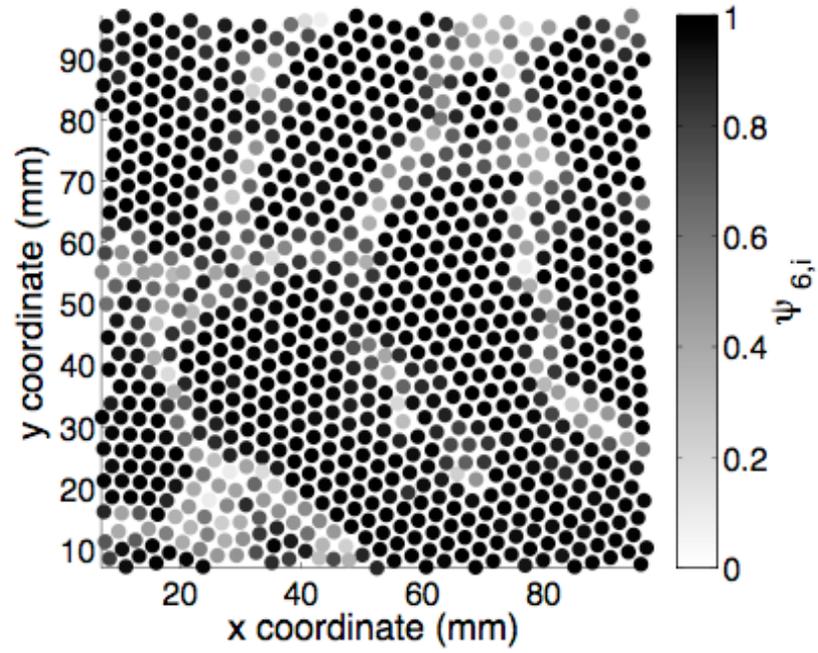 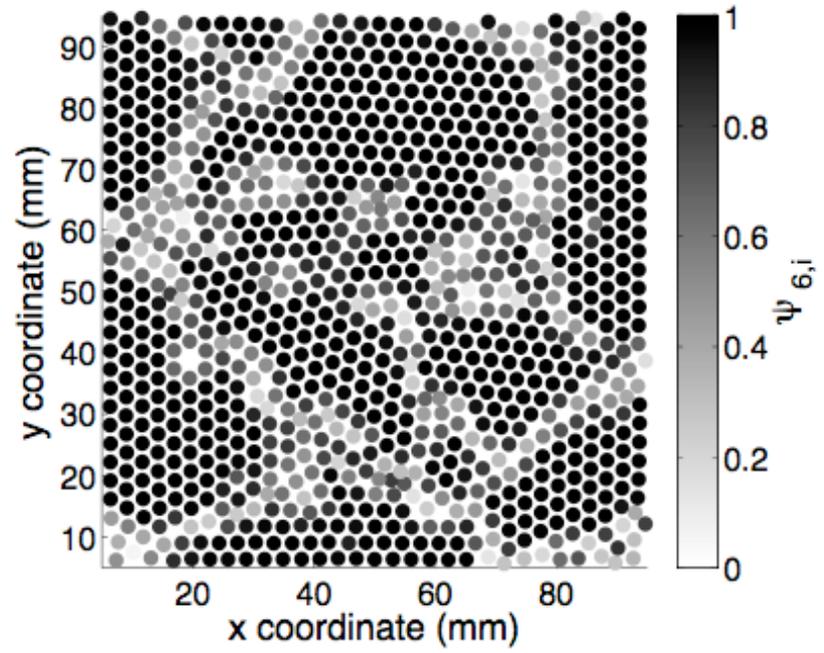

**(a)** **(b)**

FIGURE 4

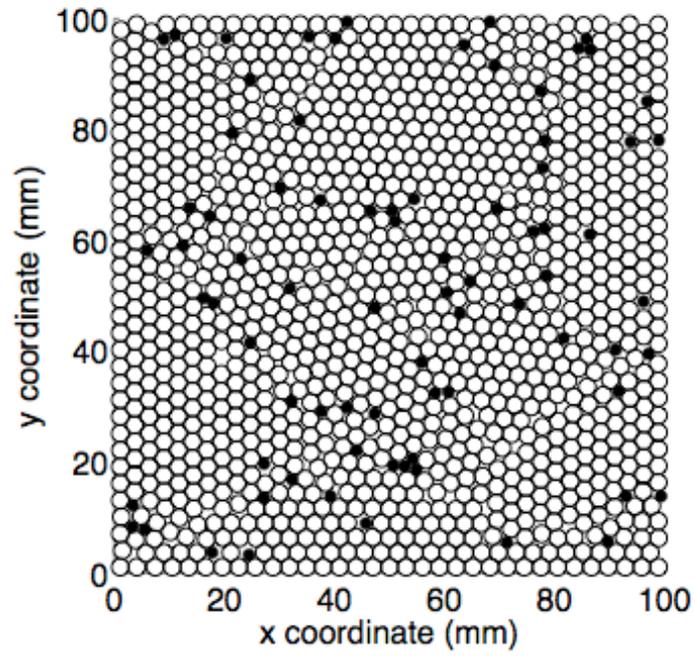

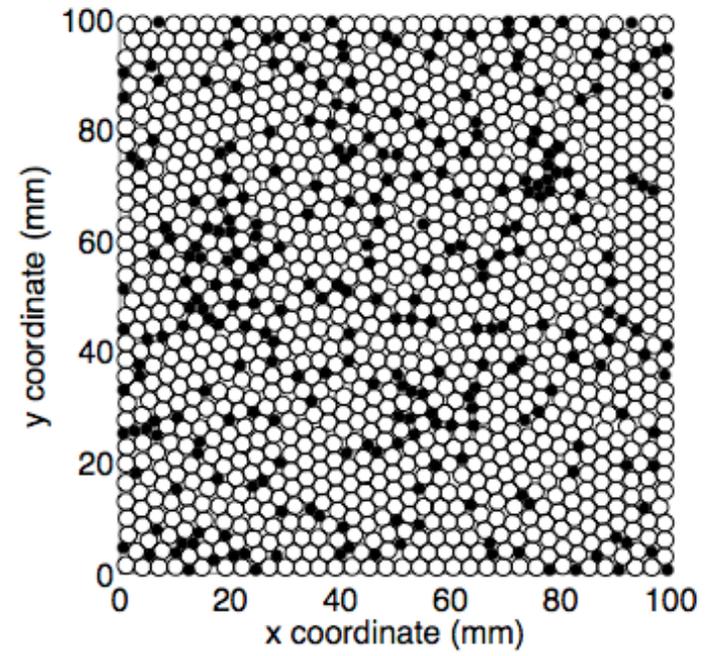

**(a)**                                                                                                      **(b)**

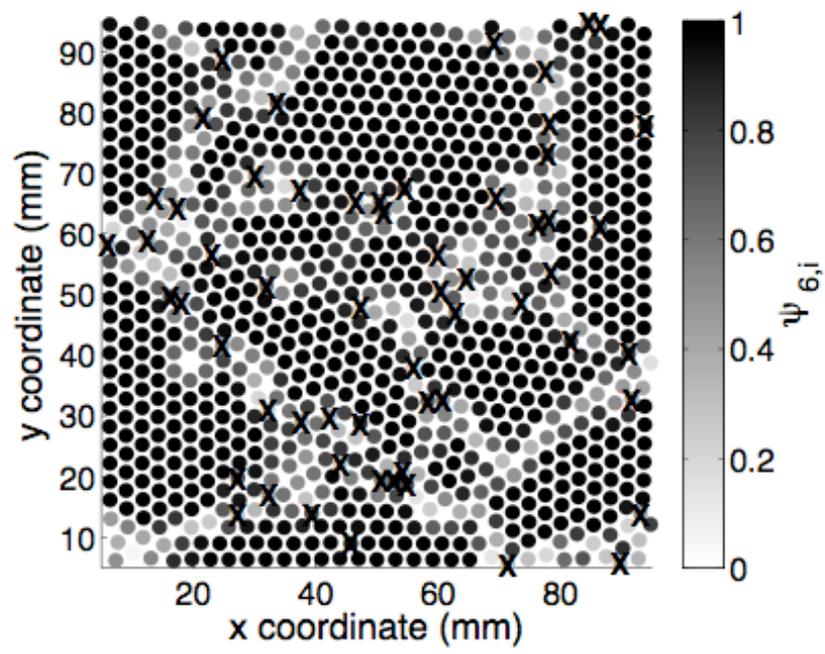

FIGURE 6

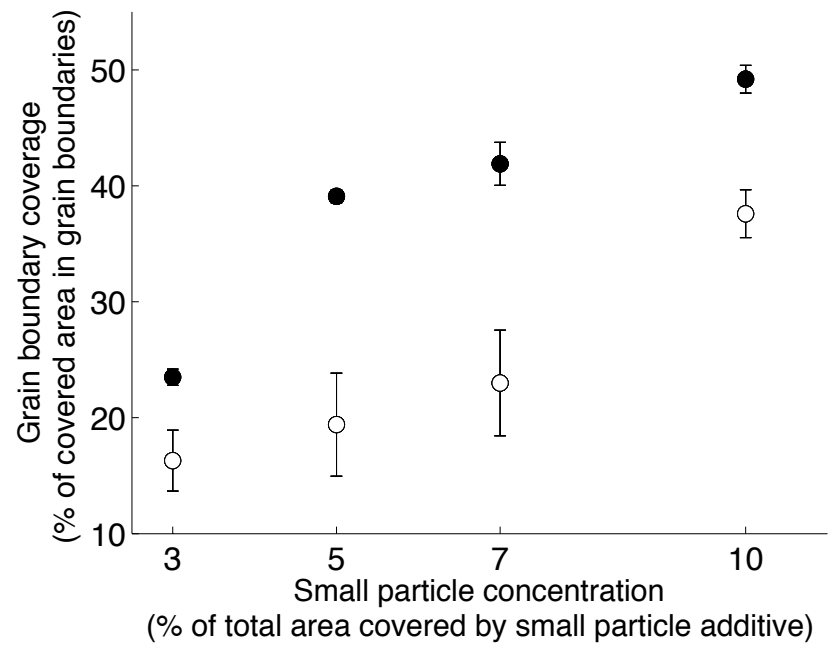

FIGURE 7

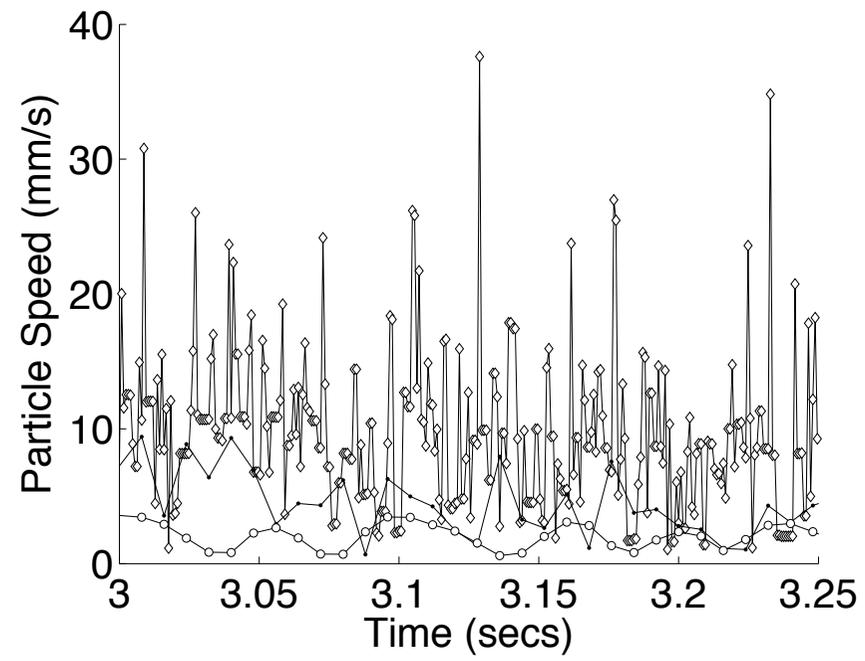

FIGURE 8

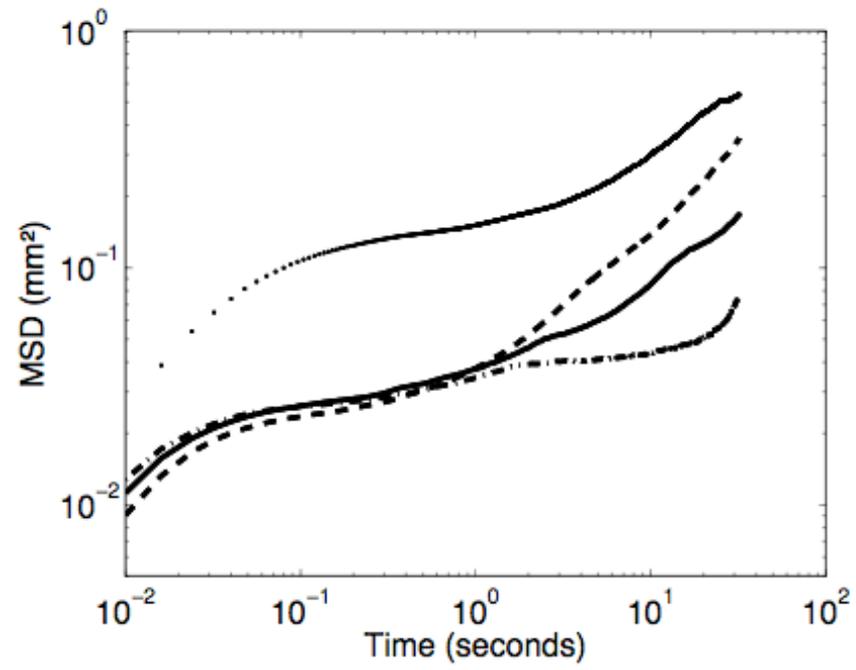

FIGURE 9

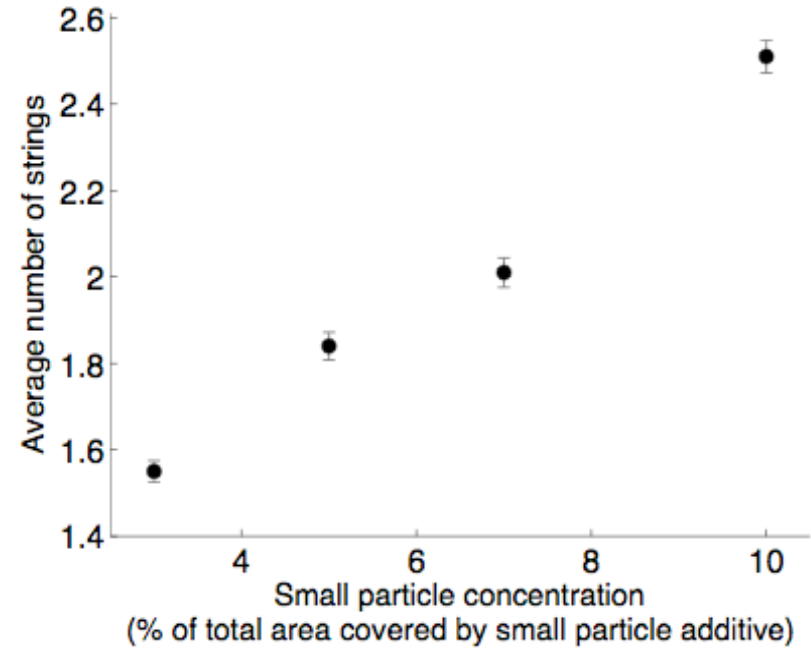

FIGURE 10

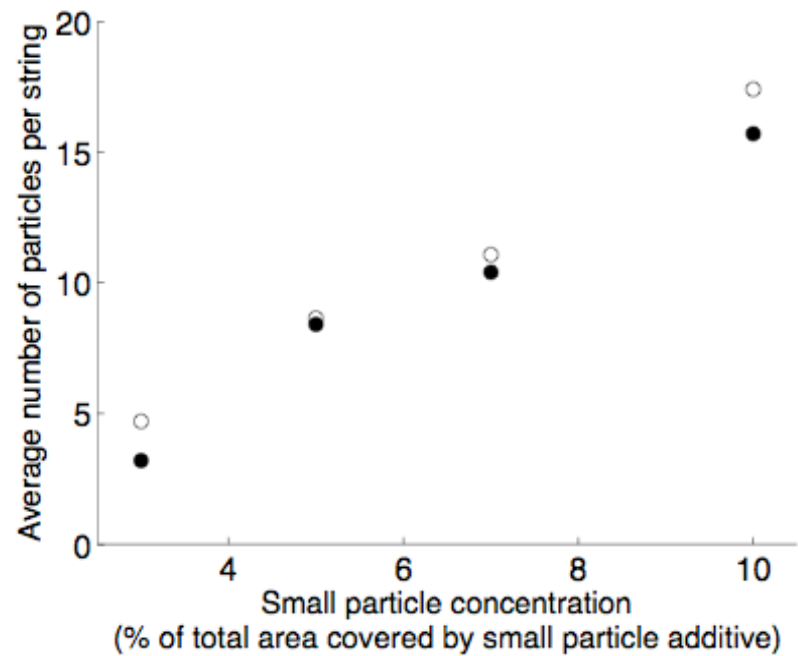

FIGURE 11

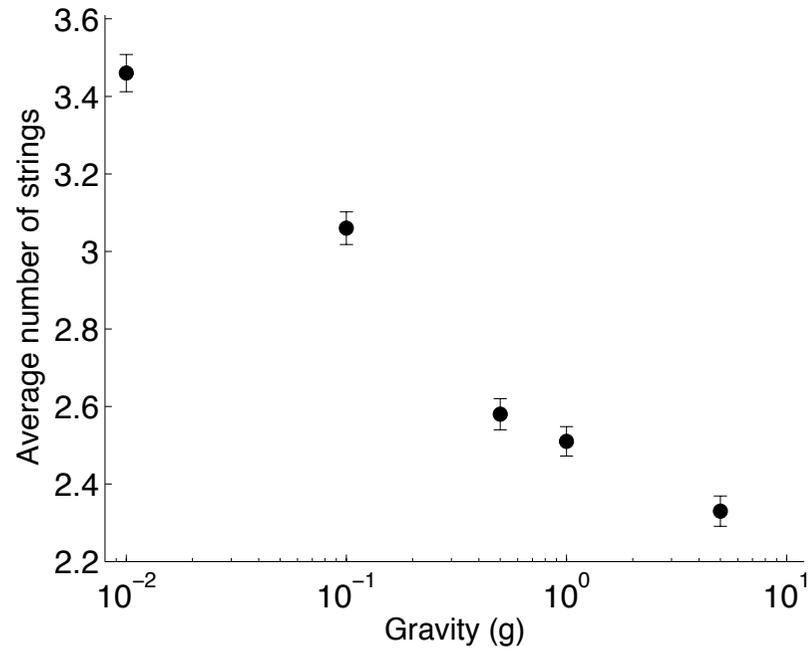

(a)

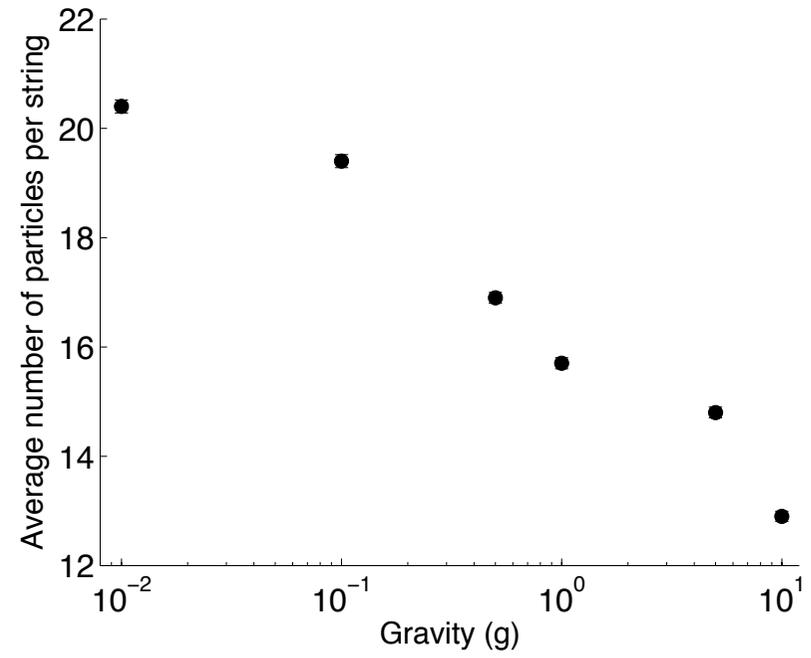

(b)

FIGURE 12

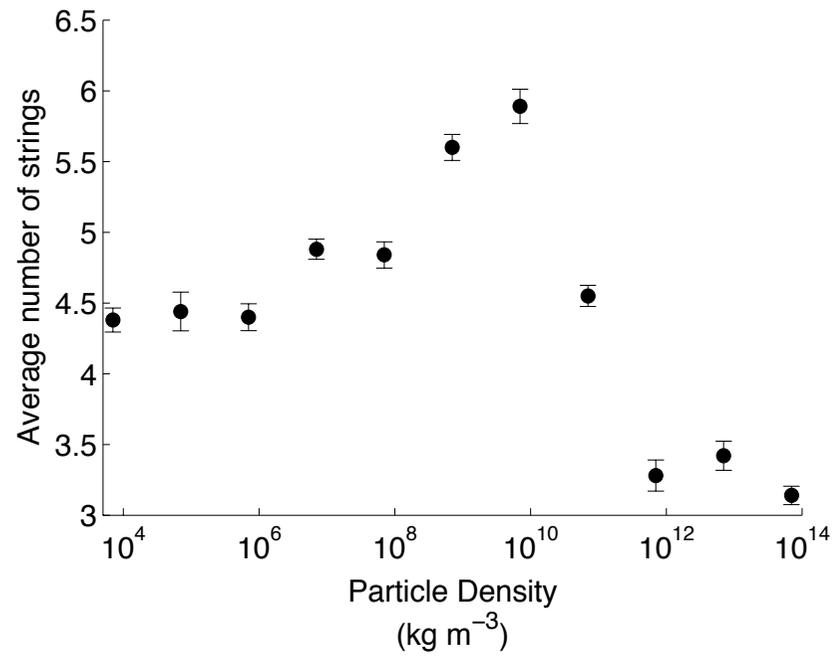

(a)

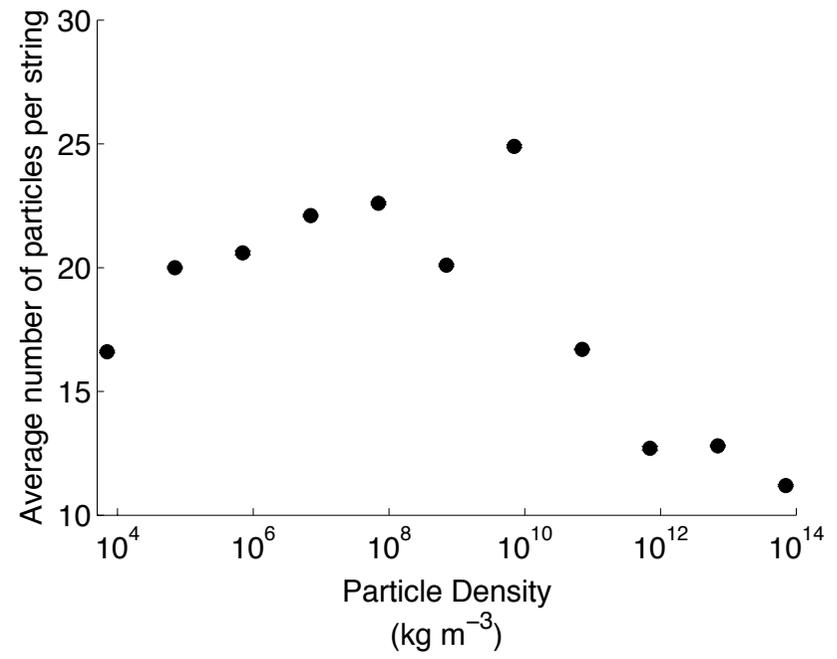

(b)